\documentstyle[psfig,12pt]{article}
\def\VEV#1{\left\langle #1\right\rangle}
\def\be{\begin{equation}}       
\def\ee{\end{equation}}
\def\bear{\be\begin{array}}      
\def\eear{\end{array}\ee}
\def\bea{\begin{eqnarray}}
\def\eea{\end{eqnarray}}

\include{mat}
\def\21{$SU(2) \ot U(1)$}
\def\ot{\otimes}

\def\bold#1{\setbox0=\hbox{$#1$}
     \kern-.025em\copy0\kern-\wd0
     \kern.05em\copy0\kern-\wd0
     \kern-.025em\raise.0433em\box0 }
\begin{document}
\begin{titlepage}
\begin{flushright}
IFIC/97-93\\
FTUV/97-63\\
hep-ph/9712213\\
November 1997
\end{flushright}
\vspace*{5mm}
\begin{center} 
{\Large \bf R--Parity Breaking in Minimal Supergravity}\\[15mm]
{\large Marco Aurelio D\'\i az} \\
\hspace{3cm}\\
{\small Departamento de F\'\i sica Te\'orica, IFIC-CSIC, Universidad de Valencia}\\ 
{\small Burjassot, Valencia 46100, Spain}
\end{center}
\vspace{5mm}
\begin{abstract}

We consider the Minimal Supergravity Model with universality of scalar
and gaugino masses plus an extra bilinear term in the superpotential
which breaks R--Parity and lepton number.  We explicitly check the
consistency of this model with the radiative breaking of the
electroweak symmetry. A neutrino mass is radiatively induced, and 
large Higgs--Lepton mixings are compatible with its experimental bound. 
We also study briefly the lightest Higgs mass. This one--parameter 
extension of SUGRA--MSSM is the simplest way of introducing R--parity 
violation.

\end{abstract}

\vskip 5.cm
\noindent ${}^{\dag}$ Talk given at the International Europhysics 
Conference on High Energy Physics, EPS--HEP--1997, 19--26 August 1997,
Jerusalem.

\end{titlepage}

\setcounter{page}{1}

The Minimal Supersymmetric Standard Model (MSSM) \cite{MSSMrep}
contains a large number
of soft supersymmetry breaking mass parameters introduced explicitly
in order to break supersymmetry without introducing quadratic 
divergencies. When the MSSM is embedded into a supergravity inspired
model (MSSM--SUGRA), the number of unknown parameters can be greatly
reduced with the assumption of universality of soft parameters at the
unification scale. In addition, in MSSM--SUGRA the breaking of the
electroweak symmetry can be achieved radiatively due to the large
value of the top quark Yukawa coupling.

The most general extension of the MSSM which violates $R$--parity 
\cite{HallSuzuki} contains almost 50 new parameters, all of them 
arbitrary although constrained by, for example, proton stability . 
The large amount of
free parameters makes R--parity violating scenarios less attractive.
Nevertheless, models of spontaneous R--parity breaking do not include
trilinear R--parity violating couplings, and these models only generate
bilinear R--parity violating terms \cite{MV_RIV}. 

Motivated by the spontaneous breaking of R--parity, we consider here 
a model where a bilinear R--parity violating term of the form
$\epsilon_i\widehat L_i^a\widehat H_2^b$ is introduced explicitly 
in the superpotential \cite{RPotros}. We demonstrate that this 
``$\epsilon$--model''
can be successfully embedded into supergravity, i.e., it is compatible
with universality of soft mass parameters at the unification scale
and with the radiative breaking of the electroweak group \cite{epsrad}. 

For simplicity we consider that only the third generation of leptons
couples to the Higgs. Therefore, our superpotential is
\begin{equation} 
W=\varepsilon_{ab}\left[
 h_t\widehat Q_3^a\widehat U_3\widehat H_2^b
+h_b\widehat Q_3^b\widehat D_3\widehat H_1^a
+h_{\tau}\widehat L_3^b\widehat R_3\widehat H_1^a
-\mu\widehat H_1^a\widehat H_2^b
+\epsilon_3\widehat L_3^a\widehat H_2^b\right]
\label{eq:Wsuppot}
\end{equation}
where the last term is the only one not present in the MSSM. This term 
induces a non--zero vacuum expectation value of the tau sneutrino,
which we denote by $\VEV{\tilde\nu_{\tau}}=v_3/\sqrt{2}$.

The $\epsilon_3$--term cannot be rotated away by the redefinition 
of the fields 
\begin{equation}
\widehat H_1'={{\mu\widehat H_1-\epsilon_3\widehat L_3}\over{
\sqrt{\mu^2+\epsilon_3^2}}}\,,\qquad
\widehat L_3'={{\epsilon_3\widehat H_1+\mu\widehat L_3}\over{
\sqrt{\mu^2+\epsilon_3^2}}}\,,
\label{eq:rotation}
\end{equation}
and in this sense the $\epsilon_3$--term is physical. If the previous 
rotation is performed, the bilinear R--Parity violating term disappear 
from the superpotential. Nevertheless, a trilinear R--Parity violating 
term is reintroduced in the Yukawa sector and it is proportional to 
the bottom quark Yukawa coupling. In addition, bilinear terms which
induce a non--zero vacuum expectation value of the tau sneutrino
reappear in the soft terms, and therefore, the vacuum expectation
value of the tau sneutrino is also non--zero in the new basis:
$\VEV{\tilde\nu'_{\tau}}=v'_3\neq 0$. These terms are
\begin{equation}
V_{soft}=(B_2-B){{\epsilon_3\mu}\over{\mu'}}\widetilde L'_3
H_2+(m_{H_1}^2-M_{L_3}^2){{\epsilon_3\mu}\over{\mu'^2}}
\widetilde L'_3H'_1+h.c.+...
\label{SoftRotated}
\end{equation}
where $\mu'^2=\mu^2+\epsilon_3^2$, $B$ and $B_2$ are the bilinear
soft breaking terms associated to the next-to-last and last terms in
eq.~(\ref{eq:Wsuppot}), and $m_{H_1}$ and $M_{L_3}$ are the soft mass
terms associated to $H_1$ and $\widetilde L_3$.

The presence of the $\epsilon_3$ term and of a non--zero vev of the
tau sneutrino induce a mixing between the neutralinos and the tau neutrino.
As a consequence, a tau neutrino mass is generated which satisfy
$m_{\nu_{\tau}}\sim (\epsilon_3v_1+\mu v_3)^2$. The quantity
inside the brackets is proportional to $v'_3$, thus a non--zero vev of the
tau sneutrino in the rotated basis is crucial for the generation of
a mass for the tau neutrino.

\begin{figure}
\centerline{\protect\hbox{\psfig{file=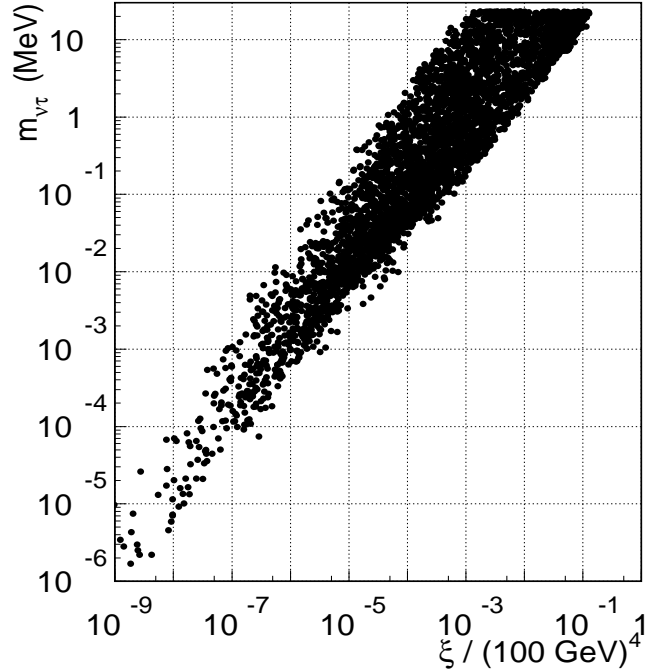,height=9.5cm,width=0.56\textwidth}}}
\caption{Tau neutrino mass as a function of  
$\xi\equiv(\epsilon_3v_1+\mu v_3)^2$, which is related to the v.e.v. of the
tau sneutrino in the rotated basis through $\xi=(\mu'v_3')^2$.}
\label{mnt_xi_ev}
\end{figure}

We assume at the unification scale universality of soft scalar masses, 
gaugino masses, soft bilinear mass parameters, and soft trilinear mass 
parameters. Using the RGE's given in \cite{epsrad} we impose the correct
electroweak symmetry breaking. In order to do that, we impose that the
one--loop tadpole equations are zero, and find the three vacuum expectation
values. This tadpole method is equivalent to use the one--loop effective
potential \cite{diazhaberii}. The solutions we find are displayed as
scatter plots. In Fig.~\ref{mnt_xi_ev} we have the induced tau neutrino mass
$m_{\nu_{\tau}}$ as a function of the combination
$\xi\equiv(\epsilon_3v_1+\mu v_3)^2$, which is related to the v.e.v.
of the tau sneutrino in the rotated basis through $\xi=(\mu'v_3')^2$.

In Fig.~\ref{mnt_xi_ev} we see that we find plenty of solutions
with values of the tau neutrino mass compatible with experimental bounds.
The reason is that in models with universality of soft supersymmetry
breaking parameters it is natural to find small values of the v.e.v.
$v'_3\sim(\epsilon_3v_1+\mu v_3)$. This can be understood if we look at
the tree level tadpole corresponding to the tau sneutrino in the rotated
basis. The relevant linear term is $V_{linear}=t'_3\tilde\nu'^R_{\tau}+...$,
with $\tilde\nu'^R_{\tau}=\sqrt{2}Re(\tilde\nu'_{\tau})-v'_3$, and the 
tree level tadpole equation is
\begin{eqnarray}
t'_3&=&(m_{H_1}^2-M_{L_3}^2){{\epsilon_3\mu}\over{\mu'^2}}v'_1
+(B_2-B){{\epsilon_3\mu}\over{\mu'}}v'_2
+{{m_{H_1}^2\epsilon_3^2+M_{L_3}^2\mu^2}\over{\mu'^2}}v'_3
\nonumber\\&&
+{\textstyle{1\over8}}(g^2+g'^2)v'_3(v'^2_1-v_2^2+v'^2_3)=0
\label{tadpoleiii}
\end{eqnarray}
It is clear that the first two terms are generated radiatively, because
at the unification scale we have $m_{H_1}=M_{L_3}$ and $B_2=B$.
The RGE's of these parameters are such that at the weak scale we
have non--zero differences $(m_{H_1}^2-M_{L_3}^2)$ and $(B_2-B)$ 
generated at one--loop and proportional to $h_b^2/(16\pi^2)$, where 
$h_b$ is the bottom quark Yukawa coupling. If for a moment we neglect these
radiative corrections, the first two terms in eq.~(\ref{tadpoleiii})
are zero and as a consequence $v'_3=0$, implying that the induced tau
neutrino mass is zero. In reality this is not the case, and the tau neutrino 
mass is radiatively generated \cite{Basis}. 

\begin{figure}
\centerline{\protect\hbox{\psfig{file=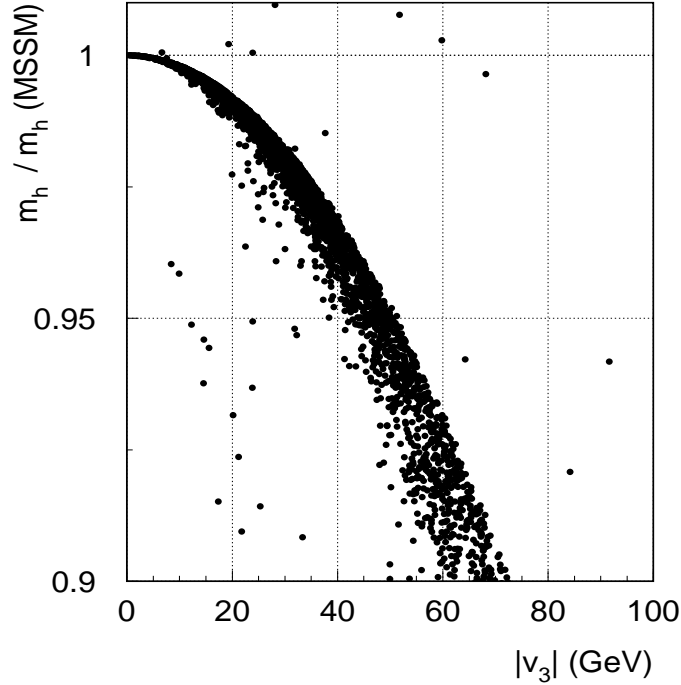,height=9.5cm,width=0.56\textwidth}}}
\caption{Ratio between the lightest CP-even neutral scalar mass in the 
$\epsilon$--model and the lightest CP--even Higgs mass in the MSSM, as a 
function of the tau sneutrino vacuum expectation value $v_3$.}
\label{ratio_v3}
\end{figure}
In this model, the CP--even Higgs bosons of the MSSM mix with the real
part of the tau sneutrino \cite{phenom}. For this reason, the neutral
CP--even scalar sector contains three fields and the mass of the lightest
scalar is different compared with the lightest CP--even Higgs of the MSSM.
In Fig.~\ref{ratio_v3} we plot the ratio between the mass of the lightest
CP--even neutral scalar in the $\epsilon$--model and the lightest
CP--even Higgs of the MSSM, as a function of the v.e.v. of the tau
sneutrino in the unrotated basis. In the radiative corrections to these 
masses we have included the most important contribution which is
proportional to $m_t^4$. As it should, the ratio approaches to unity
as $v_3$ goes to zero. Most of the time the effect of $v_3$ is to reduce 
the scalar mass, but there are a few points where the opposite happens.

In summary, we have proved that a bilinear R--Parity violating term 
can be successfully embedded into supergravity, with universality of 
soft mass terms at the unification scale, and with radiative breaking
of the electroweak symmetry. In addition, the induced neutrino mass 
is generated radiatively at one--loop, and therefore it is 
naturally small. This is a one parameter ($\epsilon_3$) extension of
MSSM-SUGRA, and therefore the simplest way to study systematically 
R--Parity violating phenomena.

\section*{Acknowledgments}

The author is indebted to his collaborators A. Akeroyd, 
J. Ferrandis, M.A. Garcia--Jare\~no, A. Joshipura, J.C. Rom\~ao, 
and J.W.F. Valle for their contribution to the work presented here.
The author was supported by a postdoctoral grant from Ministerio de 
Educaci\'on y Ciencias, by DGICYT grant PB95-1077 and by the EEC 
under the TMR contract ERBFMRX-CT96-0090.

%

\end{document}